\documentclass[12pt]{article}
\usepackage{epsf}
\usepackage{graphicx}

\begin{document}
\noindent
{\center{\bf  { Discrete  Breather and Soliton-Mode Collective
  Excitations in Bose-Einstein Condensates in a Deep Optical Lattice with Tunable Three-body Interactions }}\\}
{\center
 Galal Ahmed Alakhaly and Bishwajyoti Dey\\
             Department of Physics, University of Pune,\\
             Pune-411007, India.\\}
\vspace*{0.3in}

\noindent
{\bf Abstract}: We have studied the dynamic evolution of  the collective excitations in Bose-Einstein condensates in a deep optical lattice with tunable three-body interactions.   Their dynamics is governed by a high order discrete nonlinear Schrodinger equation (DNLSE).  The dynamical phase diagram of the system is obtained using the variational method. The dynamical evolution shows very  interesting features. The  discrete breather phase totally disappears in the regime where the three-body interaction completely dominates over the two-body interaction. The soliton phase  in this particular regime exists only when the soliton line approaches  the critical line in the phase diagram. When weak two-body interactions are reintroduced into this regime, the discrete breather solutions reappear, but occupies a very small domain in the phase space. Likewise, in this regime, 
the soliton as well as the discrete breather phases completely disappear if the signs of the two-and three-body interactions are opposite.  We have analysed the causes of this unusual dynamical evolution of the collective excitations of the Bose-Einstein condensate with tunable interactions. We have also performed direct numerical simulations of the governing DNLS equation to show the existence of the discrete soliton solution as predicted by the variational calculations, and also to check the long term stability of the soliton solution. \\

\noindent
{\bf 1. Introduction}\\

The unprecedented control over ultra-cold atoms in optical
lattices allows the Bose-Einstein condensates (BECs) formed by such atoms 
to model various physical phenomena.  A few examples are Bloch oscillation, superfluid to Mott-insulator
transition, Efimov physics etc. Efimov state were first experimentally 
observed in BECs \cite{kraemer} 35 years after its prediction by Efimov
 \cite{efimov}. In the Efimov state of the condensate, the three-body
 interaction can completely dominate over the two-body
 interaction. For a very recent paper on the Efimov effect in BECs see \cite{huang}.  Effective multi-body interactions has also been
experimentally observed in BECs \cite{will}. It is well known that the Feshbach
resonance mechanism allows fine tuning of the two-body interactions in
BECs.  In a recent paper Safavi-Naini et. al. \cite{safavi} proposed an interesting mechanism for tuning the onsite  three-body  interaction using Efimov states
\cite{efimov,ferlaino1} and its radio frequency (RF) coupling with the triply occupied state.  They found that a strong three-body attraction affects the order of the Mott-insulator-superfluid transition. Such RF coupling of the trimer state
 has also been experimentally observed in ultra cold Fermi gases
 \cite{lompe}. Very recently, Daley and Simon suggested an experimental method to make coherent three-body interactions dominate the physics of ultracold lattice gases \cite{daley}.

BEC exhibit a range of many-body phenomena, such as collective excitations
which are significantly affected by the inter-atomic interactions. 
  Soliton and breather modes are examples of such collective excitations. These nonlinear modes are localised in space,  and arises due to the balance between nonlinearity and dispersion in the system \cite{stringari}. Such localised collective modes has been experimentally observed in BECs \cite{eiermann}.
These nonlinear collective modes  also occurs in nonlinear lattice systems and are known as discrete breathers and discrete solitons. Nonlinear lattices are used to model energy localisation and transport in physics and biology \cite{dey}.  Collective excitations  modes of a fermionic gas of $^6$Li atoms in the BEC-BCS crossover regime has also been experimentally observed \cite{bartenstein} and it shows a critical transition
with variation of the strength of  inter-atomic interactions. Analytical studies of the effects of two- and three-body inteactions, and also the anharmonicity of the trap potential  on the stability and  collective excitations of BECs,  are also reported in the literature \cite{yong}. However not much is known about the localised collective excitation of BECs loaded in an optical lattice. 
It is natural to ask how the two- and three-body interactions affects the nonlinear localised  collective excitations of BECs in a deep optical lattice.

In this paper, we study how the dynamical evolution of the collective
excitations in BEC is affected by the presence of local and tunable
three-body interactions.  For this we obtain the dynamical phase
diagram of the system using the time-dependent variational method. We find that the  dynamical evolution of the collective excitations of the condensate in the regime where three-body interaction dominates over the two-body interaction shows very unusual and interesting behaviour. We have also performed direct numerical simulations of the governing equation of the system to show the existence of the discrete soliton solutions as predicted by the variational calculations, and also to check the long term stability of such solutions. 

The plan of the paper is as follows: In Section 2, we  derive 
 the higher order discrete nonlinear Schrodinger
equation (DNLS), which describes the dynamics of the scalar BECs in a deep optical lattice. In Section 3, we  describe the variational dynamics to study the
phase diagram of the system. In Section 4, we  report various phases of the system as obtained from the variational analysis and plot the phase diagrams. In Section 5, we report results of the direct numerical simulations of the higher order 
DNLS equation.  Finally we give a conclusion in Section 6.\\

\noindent
{\bf 2. Model}\\

The dynamics of the BEC with two- and three-body interactions is given by the effective mean-field 1D {\it quintic} time dependent Gross-Pitaevskii (GP) equation or the nonlinear Schrodinger equation (NLS) 
\begin{equation}
i\hbar \frac{\partial \Phi}{\partial t} = -\frac{\hbar^2}{2m}\nabla ^2 \Phi + [v_{ext} + g_0 \mid \Phi\mid^2 + g_2 \mid \Phi\mid^4]\Phi 
\end{equation}
where $g_0$ and $g_2$ are the coupling constants for the two- and three-body interactions which depend on the s-wave scattering length $a$ and the atomic mass.  The condensate is trapped in an optical lattice potential given by 
 $v_{ext} = U_l (x,y)\sin^2(2\pi z/\lambda)$  
where $\lambda$ is the wave-length of the laser, $\lambda/2$ is the lattice spacing of the optical lattice and $U_l(x,y)$ is determined by the intensity of the laser beam.
Eq. (1) has been derived  from various approaches such as
renormalisation group approach \cite{kolo},  from many-body
Schrodinger equation \cite{lieb} and also from microscopic theory
\cite{kohler}. For the deep optical lattice (for large enough laser
power) one can  use the tight-binding approximation, where the strongly
localised wave functions (atomic orbitals) at the well of the lattice can be
approximated by the Wannier functions. Accordingly, we write the condensate order parameter $\Phi({\bf r}, t)$ as a linear combination of the atomic orbitals (LCAO) $\phi({\bf r} - {\bf r}_n) = \phi_n({\bf r})$ localised in the trap $n$ 
as  
\begin{equation}
\Phi({\bf r}, t) = \sqrt {N_T}\sum_n \psi_n(t)\phi({\bf r} - {\bf r}_n)
\end{equation}
where $N_T$ is the total number of atoms. The orthonormal conditions
of the atomic orbitals (Wannier functions) gives
\begin{eqnarray}
 \int d{\bf r}\phi_n({\bf r})\phi_{n+1}({\bf r}) &=& 0  \nonumber \\ 
 \int d{\bf r}\phi_n^2({\bf r}) &=& 1   
\end{eqnarray}
\noindent
where $\psi_n(t)$ is the $n$-th amplitude which can be written in terms of the number of particles $N_n$ and phases $\theta_n$ in trap $n$ as
\begin{equation}
\psi_n(t) = \sqrt {\rho_n (t)} e^{i\theta_n (t)} 
\end{equation}
where $\rho_n = N_n/N_T$. For the tight-binding approximation (Eq. (2)) to
be valid, the depth of the optical lattice should be much larger than
the chemical potential $\mu$ and the energy of the system should be
confined to the lowest band \cite{kittel}. Substituting Eq. (2) in
equation Eq. (1), the GP equation reduces to the
discrete nonlinear Schrodinger equation (DNLSE) with higher order nonlinearity 
\cite{andrea, fatka, alfimov}
\begin{equation}
i\frac{\partial \psi_n}{\partial t} = -\frac{1}{2}(\psi_{n-1} + \psi_{n+1}) + (\epsilon_n + \Lambda_0\mid\psi_n\mid^2 + \Lambda_2\mid\psi_n\mid^4)\psi_n 
\end{equation}
where  
\begin{equation} 
\Lambda_0 = \frac{g_0N_T}{2K}\int d{\bf r} [\phi^4_n], 
\end{equation}
\begin{equation}
\Lambda_2 =  \frac{g_2N_T^2}{2K}\int d{\bf r} [\phi^6_n] 
\end{equation}
and 
 \begin{equation} 
\epsilon_n = \frac{1}{2K}\int d{\bf r}[\frac{\hbar^2}{2m}(\nabla \phi_n)^2
 + v_{ext} \phi_n^2]. 
\end{equation}
Here time has been rescaled as $t \rightarrow \frac{\hbar}{2K} t$, where
\begin{equation}
K =  -\int d{\bf r}[\frac{\hbar^2}{2m}\nabla \phi_n \cdot \nabla \phi_{n\pm 1} + \phi_n v_{ext} \phi_{n\pm 1}], 
\end{equation}
Eq. (5) can be obtained from the Hamiltonian equation of motion 
\begin{equation}
\dot \psi_n = \frac {\partial {\it H}}{\partial (i\psi_n^*)}
\end{equation}
 where the Hamiltonian function is given by
\begin{equation}
 H = \sum_n [-\frac{1}{2} (\psi_n\psi_{n+1}^* + \psi_n^* \psi_{n+1}) +
 \epsilon_n \mid \psi_n\mid^2 + \frac{\Lambda_0}{2}\mid \psi_n\mid^4 +
 \frac{\Lambda_2}{3}\mid \psi_n\mid^6]
\end{equation}
with $i\psi_n^*$ and $\psi_n$ as conjugate variables.
Both the Hamiltonian and the norm $\sum_n\mid \psi_n\mid^2 = 1$ are
conserved.\\

\noindent
{\bf 3. Variational Dynamics}\\

We study the dynamics of the BEC using the time-dependent variational method. The basic idea is to take a trial function which depends on a number of time-dependent parameters, and to derive equations of motion for these parameters using the variational principle. The amplitude of the wavefunction determines the density distribution, while its phase determines the velocity field. As mentioned above, the interaction between the particles leads to collective behaviour of the many particle system. It is therefore useful to write the variational wavefunction in terms of variational parameters, such as a center-of-mass coordinate and the width of the condensate, hence describing the collective behaviour of the condensate \cite{pethick}.  For this, we study the dynamical evolution of a Gaussian profile wave packet 
\cite{andrea, ripoli, zoller} 
\begin{equation}
\psi^{n}_{V}(t)=\sqrt{k}.exp\left\{
  -\frac{(n-\zeta)^{2}}{\gamma^{2}}+
  ip(n-\zeta)+i\frac{\delta}{2}(n-\zeta)^{2}\right\}
\end{equation}
 where the
variational parameters $\zeta(t)$ and $\gamma(t)$ are the center and
the width respectively of the density $\rho_n = \mid\psi_n\mid^2$, $p(t)$ and $\delta(t)$ are their associated momenta, and $k$ is a normalisation factor. The linear site dependence of the phase term shows that the velocity or the conjugate momentum of the center-of-mass coordinate $\zeta (t)$ is site independent. Similarly the  quadratic site dependence of the phase term shows that the velocity or the conjugate momentum of the width $\gamma (t)$ depends linearly on site $n$ \cite{pethick}. The dynamical
evolution of the variational wavepacket can be obtained  by a
variational on  the Lagrangian 
\begin{equation}
 L = \sum_{n=-\infty}^{n=\infty} i\dot\psi_n\psi_n^* -  H
\end{equation}
Using Eqs. (11) and (12), and  after rearrangement, $L$ can be written as  
\begin{equation}
 L = p\dot{\zeta}-\frac{\gamma^{2}\dot{\delta}}{8}-\left[\frac{\Lambda_{0}}{2\sqrt{\pi}\gamma}\right]- \left[\frac{2\Lambda_{2}}{3\sqrt{3}\pi\gamma^{2}}\right] +\cos{p} e^{-\eta}-V(\gamma,\zeta)
\end{equation}
where 
\begin{equation}
\eta = \frac{1}{2\gamma^2} + \frac{\gamma^2\delta^2}{8} 
\end{equation}
 and
\begin{equation}
V= k\int dn [\epsilon_n e^{{-2(n-\zeta)^2}/\gamma^2}]
\end{equation}
The equation of motion for the variational parameters $q_{i}$, where $q_{i}=p$, $\zeta$,  $\gamma $ and $\delta $,  can be obtained 
from the Euler-Lagrange equations 
\begin{equation}
 \frac{d}{dt}\left( \frac{\partial L}{\partial \dot{q_{i}}}
\right)=\frac{\partial L}{\partial q_{i}}
\end{equation}
  as
\begin{equation}
\dot{p}=-\frac{\partial V}{\partial \zeta}
\end{equation}
\begin{equation}
\dot{\zeta}=\sin(p) e^{-\eta}
\end{equation}
\begin{equation}
\dot{\delta}=(\frac{4}{\gamma^{4}}-\delta^{2})\cos p e^{-\eta} +\frac{2\Lambda_{0}}{\sqrt{\pi}\gamma^{3}}+ \frac{16\Lambda_{2}} {3\sqrt{3}\pi\gamma^{4}}-\frac{4}{\gamma}\frac{\partial V}{\partial \gamma}
\end{equation}
and 
\begin{equation}
\dot{\gamma}=\gamma  \delta \cos p e^{-\eta}
\end{equation} 
The pairs ($\zeta$, $p$) and ($\frac{\gamma^{2}}{8}$, $\delta $)  are
canonically conjugate dynamical variables with respect to the
effective Hamiltonian
\begin{equation}
 H =\frac{\Lambda_{0}}{2\sqrt{\pi}\gamma}+\frac{2\Lambda_{2}}
{3\sqrt{3}\pi\gamma^{2}}-\cos p e^{-\eta} +V(\gamma,\zeta)
\end{equation}
The wave packet group velocity  is given by 
\begin{equation}
v_{g}\equiv \frac{\partial  H}{\partial p}
\end{equation}
and the inverse effective mass is given by
\begin{equation}
\frac{1}{m^{*}}=\frac{\partial^{2} H}{\partial p^{2}}=\cos p
e^{-\eta}
\end{equation}
The numerical solutions of the above coupled nonlinear ODEs (Eqs. (18-21)) along with the Hamiltonian (Eq. (22))
gives the dynamical regimes or phases of the system (Eq. (5)). Knowing the solutions, various phases can be obtained as follows:  as mentioned above, $\gamma$ and $\delta$ are conjugate variables. $\gamma$ represent the width of the wavepacket and $\delta$ denote its conjugate momenta. A plot of $\delta (t)$ versus $\gamma (t)$ gives the corresponding phase space ($\delta  -  \gamma$) trajectory. To obtain the soliton phase, we note that the solitons are shape-preserving nonlinear localised solutions. This means that the width of the soliton solution do not change with time. A soliton solution can be obtained by imposing the condition $\dot\gamma = \dot\delta = 0$. This implies that the soliton mode is be  represented by the fixed point of the $\delta  -  \gamma$ trajectory. The discrete breathers are nonlinear collective excitations which are localised in space and oscillates in time. This implies that for the discrete breather solutions, the width $\gamma (t)$ and its conjugate momenta $\delta (t)$ will oscillate in time and the corresponding $\delta -  \gamma$ trajectories will be closed curves. Similarly, for the self-trapped phase the motion of the center of the wave packet stops ($\dot\zeta \rightarrow 0$) as the effective mass tends (Eq. (24)) to infinity.  On the other hand, for the diffusive phase there is complete spreading of the wave packet and this phase corresponds to  $\gamma \rightarrow \infty$, $\dot\zeta \neq 0$, therefore the effective mass is finite. Various phases as obtained from the solutions of Eqs. (18-21) can be shown in the  phase diagram. \\

\noindent
{\bf 4. Phase diagram}\\

For a horizontal optical potential, the onsite energy $\epsilon_n$, as well as $V(\zeta, \alpha)$, are constant \cite{andrea}. Accordingly from Eq. (18) the momentum $p(t)$ is conserved. This implies $p(t) = p_0$  where $p_0$ is the initial momentum.
From Eq. (22) the conserved initial energy can be written as
\begin{equation}
H_{0}= \frac{\Lambda_{0}}{2\sqrt{\pi\alpha}}+
\frac{2\Lambda_{2}}{3\sqrt{3}\pi\alpha} -\cos p_{0}
e^{-\frac{1}{2\alpha}- \frac{\alpha \delta^{2}}{8}}
\end{equation}
For the sake of convenience we have denoted the width of the wave packet by $ \alpha$ instead of $\gamma$, where  $ \alpha=\gamma^2$.
Eq. (25) can be rewritten as
\begin{equation}
 \delta^{2}=-\frac{\left[8\alpha\log\left(\frac{\frac{ \Lambda_{0}}{2\sqrt{\pi\alpha}}+\frac{2\Lambda_{2}}{3\sqrt{3}\pi\alpha}- H_{0}} {\cos p_{0}}\right) +4 \right]} {\alpha^{2}}
\end{equation}
 The trajectories in the $ \left( \alpha - \delta \right)$ plane can be obtained from the above equation along with the solutions of the coupled equations, Eqs. (18-21). In the region $\cos p_0 > 0$, there are two phases. This can be seen as follows: since the argument of the logarithm term in Eq. (26) has to be positive, this implies that for $H_0> 0$ the width $\alpha(t)$ can attain a maximum value $\alpha_{max}$ for $t \rightarrow \infty$. The value of $\alpha_{max}$ can be obtained by equating the  argument of the logarithm term to zero. When $\alpha \rightarrow \alpha_{max}$,  we obtain from Eq. (26) that $\delta \rightarrow \infty$. Eq. (15) then implies $\eta \rightarrow \infty$ and  accordingly we obtain from Eq. (19) that  $\dot\zeta \rightarrow 0$, and  from  Eq. (24), that  $1/m^* \rightarrow 0$.  This corresponds to the self-trapped phase since the centre of the wave packet stops as the effective mass goes to infinity. Thus we see that even if the system is fully Hamiltonian (conservative), the apparent damping occurs due to the fast (exponential) growth of the effective mass with time. On the other hand, when $-\cos p_0 < H_0 < 0$,  we see from Eq. (26) that in this case
$\alpha(t \rightarrow \infty ) \rightarrow \infty$, and accordingly $\delta \rightarrow 0$ and from Eq. (15) $\eta \rightarrow 0$. But from Eq. (19) $\dot\zeta \neq 0$ and from Eqs. (24) and (25)  we see that the effective mass is finite as $ \frac{1}{m^{*}}\approx -H_{0} > 0$.  There is complete spreading of the wave packet, giving rise to the diffusive phase. The critical line which separates the self-trapped and the diffusive phases is obtained from the condition $H_0 = 0$. Substituting $H_0 = 0$ in Eq. (25), we get the critical line  (with $\delta_0 =0$)
as
\begin{equation}
\Lambda_{0, {\rm c}} = 2\sqrt{\pi \alpha_0}\left ( \cos p_0  e^{-1/2\alpha_0} - \frac{2\Lambda_2}{3\sqrt 3 \pi \alpha_0} \right )
\end{equation}
The critical line for $\cos p_0 > 0$ is shown in Fig. 1.
\begin{figure}
\includegraphics[width=180mm,height=180mm]{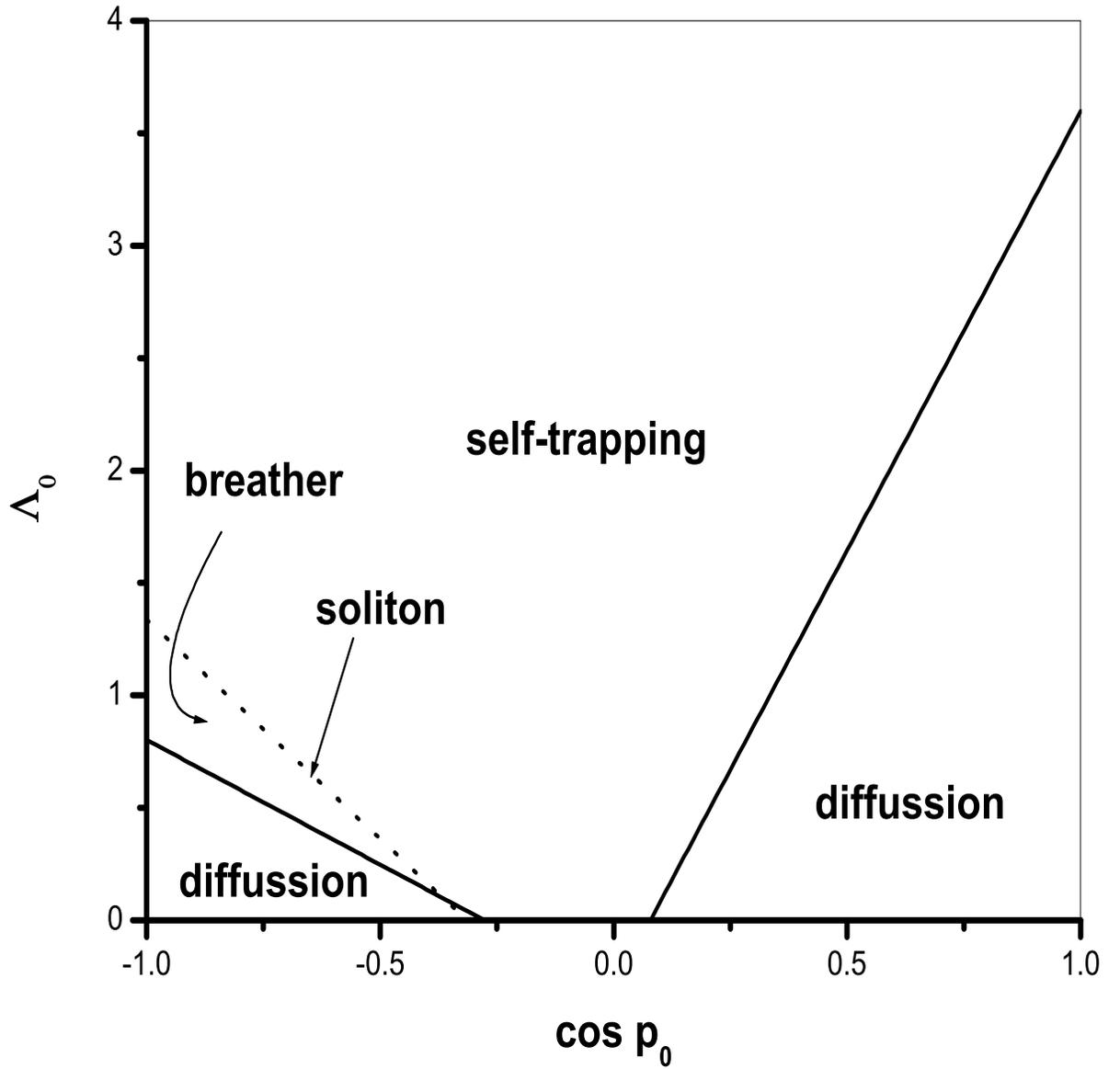}
\caption{ \label{fig.1:}Dynamical phase diagram  with two- and three-body interactions ($ \Lambda_{2}=1$) with $\alpha_{0}=2, \ \delta_{0}=0$ and $\zeta_{0}=0$.}
\end{figure}

The negative effective mass ($\cos p_0 < 0$) case is more interesting. This is because in this case, beside the self-trapping and the diffusion regimes, we get two new phases. These are the discrete soliton  and discrete breather phases. In this case, the conserved initial energy $H_0$ and the equation for the $ \left( \alpha - \delta \right)$ trajectory are same as  Eq. (25) and Eq. (26) respectively, except that  the $\cos p_0$ terms in these two equations should be replaced by $-|\cos p_0|$.
Following the same procedure as discussed above, we can see that in this case there are also self-trapping and diffusion regimes, and the 
critical line separating these two regimes is obtained from the condition $H_0 = |\cos p_0|$. Replacing $H_0$ by  $ |\cos p_0|$  in the expression of the conserved Hamiltonian, we obtain the critical line which separates the two regimes as
\begin{equation}
\Lambda_{0,  {\rm c}} = 2\sqrt{\pi \alpha_0}\left ( |\cos p_0|(1 - e^{-1/2\alpha_0}) - \frac{2\Lambda_2}{3\sqrt 3 \pi \alpha_0} \right )
\end{equation}
 As mentioned above, the soliton 
mode is represented by the fixed point of the $ \alpha -\delta $ trajectory (
$\dot{\alpha}=0$ and $\dot{\delta}=0$). Eqs. (20) and (21) then have stationary solution $\alpha(t) = \alpha_0$ and $\delta(t) = 0$ ($\alpha(t) = \gamma^2(t)$). From Eqs. (15) and (19), we obtain that  $\dot\zeta = \sin(p_0)e^{-\frac{1}{2\alpha_0}}$ = constant. This implies that for soliton solutions, the center of mass
moves with constant velocity, and the width (shape) of the soliton do not change with time. From Eq. (20) we then obtain 
\begin{equation}
\Lambda_{0, {\rm soliton}} = 2\sqrt{\frac{\pi}{\alpha_0}}\left ( |\cos p_0|e^{-1/2\alpha_0} - \frac{4\Lambda_2}{3\sqrt 3 \pi} \right )
\end{equation}
This gives the soliton line in the phase diagram. The soliton line gives the value of the two-body interaction parameter $\Lambda_0$ for which the soliton solutions are allowed. It may be noted that the soliton line corresponds to the maximum of the energy (Eq. (22)). This is because Eq. (29) can also be obtained from the condition $ \left( \frac{\partial H}{\partial \alpha}\right)_{\delta =0,\alpha_0 } = 0$  and 
for which $ \left( \frac{\partial^2 H}{\partial
 \alpha^2}\right)_{\delta =0, \alpha_0} < 0.$ 
The soliton line is shown by the dotted line in Fig. 1. Yet another interesting phase which appears in the case of the negative effective mass is  the discrete breather. Discrete
breathers are nonlinear collective  modes
which are localised in space and periodic in time. For discrete
breather solution $ \alpha (t)$ , $ \delta (t) $ and $ m^{*} $
oscillates  with time.  The center of mass travels with nearly constant velocity and with an oscillating width. Discrete breather solution appears when the nonlinearity parameter $ \Lambda_{0, {\rm c}} < \Lambda_0 < \Lambda_{0, {\rm soliton}}$. The trajectories in the $ \alpha -\delta $
plane are closed and $\alpha $ oscillates between the initial values
$\alpha_{0} $ and the maximum value $\alpha^{max}_{osc}> \alpha_{0}
$.  The complete phase diagram for the case when the two-body interactions dominates over
the three-body interactions is shown in Fig.~\ref{fig.1:}.
\begin{figure}
\includegraphics[width=200mm,height=200mm]{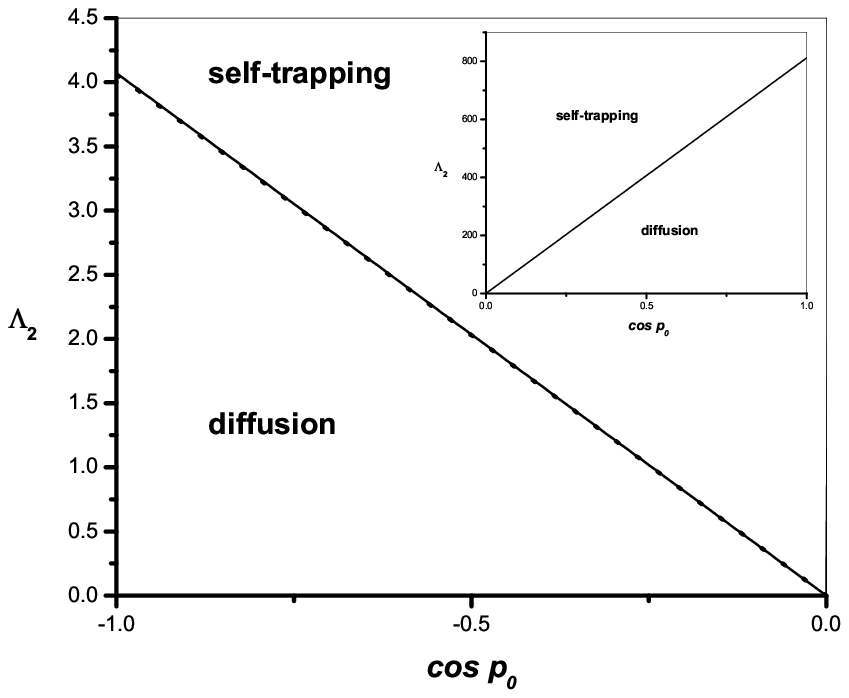}
\caption{\label{fig.2:} Dynamical phase diagram of the BEC with only three-body interaction ($ \Lambda_{2}>0$ and $ \Lambda_{0}=0$) with $\alpha_{0}=100, \delta_{0}=0$ and $\zeta_{0}=0$. The solid line represent the critical line 
($ \Lambda_{0, {\rm c}}$)  and  the  overlapping dotted line  represent the soliton line ($ \Lambda_{0, {\rm soliton}}$). The inset show the phase diagram for $\cos p_{0} >0 $.  }
\end{figure}

We can now obtain the phase diagrams for the case when the three-body interaction dominates over the two-body interaction ($\Lambda_2>>\Lambda_0$). For this, we first consider the case when the  three-body interactions completely dominate over the two-body interactions, and accordingly we set the two-body  nonlinear interaction parameter  $  \Lambda_{0}=0 $. By substituting $\Lambda_{0, {\rm c}} = 0$ in Eq. (28),  we get the critical line   $\Lambda_{2, {\rm c}}$ which separates the self-trapped  and the diffusion regimes as, 
\begin{equation}
\Lambda_{2, {\rm c}}=\frac{3\sqrt{3}\pi\alpha_{0}}{2}\left|\cos p_{0}\right| \left( 1-  e^{-1/2\alpha_{0}} \right)
\end{equation}
Similarly, by substituting $\Lambda_{0, {\rm soliton}} = 0$ in Eq. (29) we get  the  soliton line as
\begin{equation}
\Lambda_{2, {\rm soliton}}=\frac{3\sqrt{3}\pi}{4}\left|\cos p_{0}\right| e^{-1/2\alpha_{0}}
\end{equation}
\begin{figure}
\includegraphics[width=200mm,height=200mm]{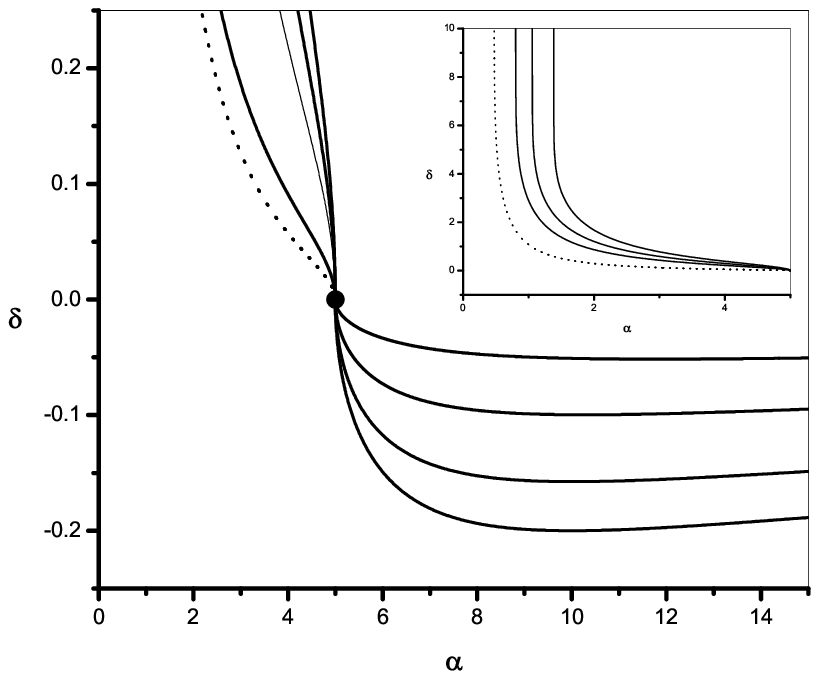}
\caption{ \label{fig.3:} $ \alpha - \delta $ trajectories  in the state  with only three-body interaction ($ \Lambda_{2}>0$ and $ \Lambda_{0}=0$). the figures are plotted for different values of $ \Lambda_{2} > 0$ and  $\cos p_{0} < 0$ with $\alpha_{0}=5$, $p_{0}= 3\pi/4 $. The  critical values of the three-body interaction parameter for different transitions are $ \Lambda_{2, {\rm c}} = 2.7462$ and $ \Lambda_{2, {\rm soliton}}= 2.61112$. Inset shows the $ \alpha - \delta $  trajectories for  $\Lambda_{2}$ very close to $\Lambda_{2, {\rm soliton}} $.}
\end{figure}

\noindent
The soliton solutions are expected when the three-body interaction
parameter $\Lambda_2$ values are on the soliton line,
i.e. $\Lambda_2=\Lambda_{2, {\rm soliton}}$.  From the above equation we can see that this can be satisfied for arbitrary value of $\alpha_0$ (initial width of the condensate density). We checked this prediction of the variational calculations by direct numerical simulations of the governing equation of the system,  the DNLS equation (Eq. (5)). However, from the direct numerical integration (Section 5) of Eq. (5) we obtain an unusual result that  the discrete soliton mode collective excitations  exist only for the large value of $\alpha_0$. From
Eqs. (30) and (31)  we see that for large value of
$\alpha_0$, $\Lambda_{2, {\rm soliton}} \rightarrow \Lambda_{2, {\rm c}}$, i.e. the
soliton solutions  exist only when the soliton line approaches the critical line in the phase diagram. Fig.~\ref{fig.2:}  shows the phase diagram for large value of $\alpha_0=100$ and we can see that in this case the soliton line (dotted line) approaches the critical line (solid line). From the direct numerical solution of the DNLS equation we find that indeed, there is a stable soliton solution for $\alpha_0=100$. Details are given in Section 5. 
 Yet another unusual result for the BEC  with only
 three-body interactions is that it does not allow discrete breather collective
 mode excitations.  The absence of the discrete breather modes in this
 regime can be seen from the following:  discrete breather collective modes occur when the three-body interaction parameter $\Lambda_{2}$ satisfies the relation $\Lambda_{2, {\rm c}} <\Lambda_{2} <\Lambda_{2, {\rm soliton}}$. As mentioned above, the trajectories in the $ \alpha -\delta $ plane are closed, and $\alpha (t) $ oscillates between the initial value $\alpha_{0} $ and the maximum value $\alpha^{{\rm max}}_{{\rm osc}}> \alpha_{0} $.  
The breather region extends until the value of parameter $\Lambda_2$ approaches  $ \Lambda_{2, {\rm breather}} $. For $ \Lambda_{2, {\rm soliton}} < \Lambda_{2} < \Lambda_{2, {\rm breather}} $, $ \alpha_{0} $ oscillates between $ \alpha_{0} $ and the minimum value $ \alpha^{{\rm min}}_{{\rm osc}} < \alpha_{0}$. 
Both the values $\alpha^{{\rm max}}_{{\rm osc}} $ and $\alpha^{{\rm min}}_{{\rm osc}} $ are roots (together with $\alpha_{0}$ ) of 
Eq. (26) for $\Lambda_0=0$ and $\cos p_0<0$. The
condition for which this equation does not have another root
$\alpha^{min}_{osc} < \alpha_{0}$  gives 
$\Lambda_{0, {\rm breather}}$ \cite{andrea}. After rearrangement, one can show that
$\Lambda_{2, {\rm breather}}$ is given by the maximum value of the r.h.s of the expression
\begin{equation}
\frac{\Lambda_{2}}{\Lambda_{2, {\rm soliton}}}= \frac{2\alpha_{0} x}{(1-x )}\left[ 1-e^{(x-1)/2\alpha_{0}x} \right]
\end{equation}
Here  $\alpha^{{\rm min}}_{{\rm osc}} = x\alpha_0$ and $x<1$. From the maximum value of the r.h.s, we find that $\Lambda_{2, {\rm breather}} \rightarrow \Lambda_{2, {\rm soliton}}$.  When $\Lambda_{2, {\rm breather}} \rightarrow \Lambda_{2, {\rm soliton}}$, the area enclosed within the $\alpha - \delta$ trajectories shrinks to zero (to the soliton fixed point) implying  nonexistence of the discrete breather modes. Also, for the three-body interaction parameter in the range $\Lambda_{2, {\rm c}} <  \Lambda_{2} <\Lambda_{2, {\rm soliton}}$ there are no closed trajectories in the $\alpha - \delta$ plane and this implies that there are no discrete breather solutions. 
This is shown in Fig.~\ref{fig.3:},  where we can see that there are no closed trajectories around the soliton fixed point. The inset of this figure shows no closed trajectories very close to $\Lambda_{2, {\rm soliton}}$.
However, if we now add a weak two-body interaction 
($\Lambda_{0}< \Lambda_{2}$), and if both the parameters $\Lambda_{0}$
and $\Lambda_{2}$ are positive, then the discrete breather solution
is allowed. In this case, after rearrangement, we obtain
\begin{figure}
\includegraphics[width=200mm,height=200mm]{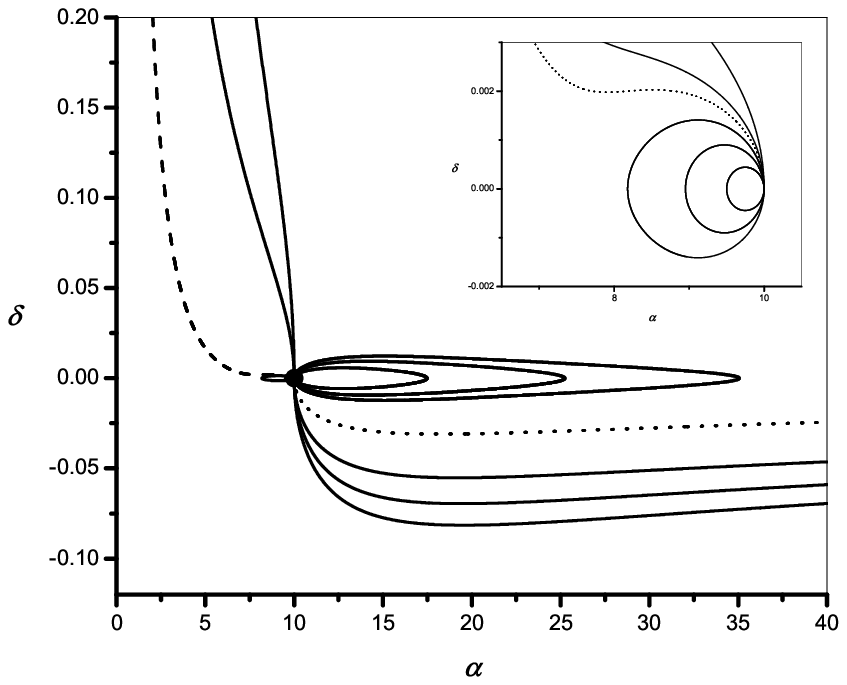}
\caption{ \label{fig.4:} $ \alpha - \delta $ trajectories in the state with weak two-body interaction ($ \Lambda_{2}>\Lambda_{0}$)
for different values of $ \Lambda_{2} > 0$ (beginning from the bottom $ \Lambda_{2}=0.5,  1,  1.5,   \Lambda_{2, {\rm c}},  2.32,  2.34,  2.36,  \Lambda_{2, {\rm soliton}},  2.382,  2.384,  \Lambda_{2, {\rm breather}}, 4, 10 )$  for the case  $\cos p_{0} < 0$ with $\alpha_{0}=5, \ p_{0}= 3\pi/4 , \Lambda_{0}=0.1 $.  Inset shows the $ \alpha - \delta $ trajectories close to $\Lambda_{2, {\rm breather}}$
(from the top $\Lambda_{2, {\rm c}} =2.39, 2.386, \Lambda_{2, {\rm breather}}, 2.384, 2.383, 2.382$).}
\end{figure}
\begin{equation}
\frac{\Lambda_{2}}{\Lambda_{2soliton}}= \frac{\frac{2\alpha_{0} x}{(1-x )}\left[ 1-e^{(x-1)/2\alpha_{0}x} \right] +\frac{ \sqrt{\alpha_{0}}\sqrt{x}\left(1-\sqrt{x}\right)}{\sqrt{\pi}(1-x)\cos p_{0} e^{-1/2\alpha_{0}}} \Lambda_{0}} {1+ \frac{\sqrt{\alpha_{0}}}{2\sqrt{\pi} \cos p_{0} e^{-1/2\alpha_{0}}}\Lambda_{0} } 
\end{equation}
\begin{figure}
\includegraphics[width=200mm,height=200mm]{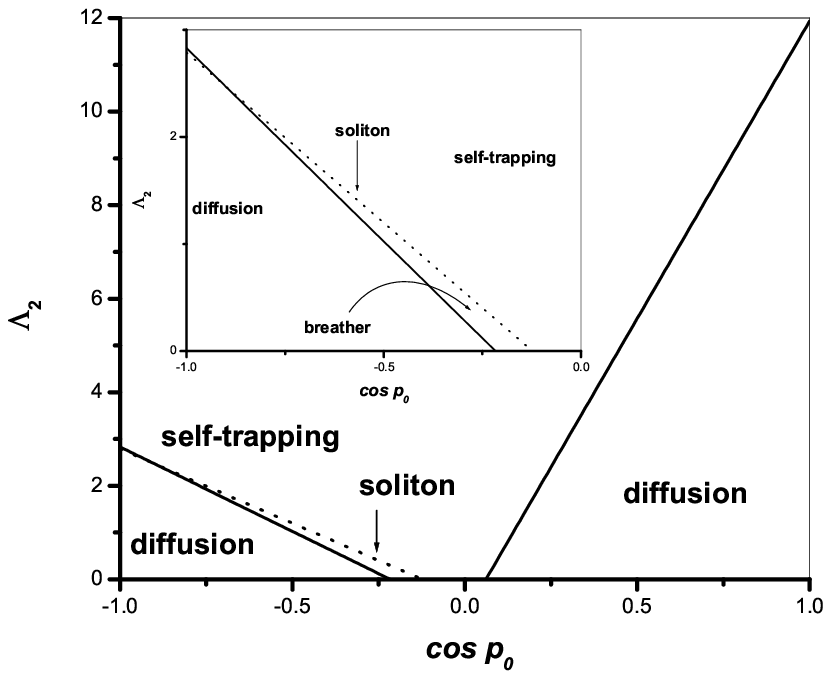}
\caption{\label{fig.5:} Dynamical phase diagram in the state  for 
  positive two- and three-body interactions  with $\alpha_{0}=2,
  \delta_{0}=0$, $\zeta_{0}=0$ and  weak two-body interaction parameter $ \Lambda_{0}=0.24$. Inset  shows the small discrete breather region.}
\end{figure}

\noindent
Again, from the maximum value of the r.h.s of Eq. (33) we find
that $\Lambda_{2, {\rm breather}} >\Lambda_{2, {\rm soliton}}$, thus satisfying the
condition for the appearance of the discrete breather solutions.  This is shown
in Fig.~\ref{fig.4:}, where we can see that there are closed trajectories (on the right of the soliton fixed point) representing the discrete breather solutions. For   $\Lambda_{2}> \Lambda_{2, {\rm soliton}}$, but very
close to $\Lambda_{2, {\rm soliton}}$, the area of the trajectories near the fixed
point is very small  but nonzero (shown in the inset of
Fig.~\ref{fig.4:} ),  thus allowing the discrete breather
solutions to exist. Fig.~\ref{fig.5:}  shows the complete phase diagram.

When $\Lambda_{0}$ have  an opposite sign to that of $\Lambda_{2}$, then  both the solutions, soliton as well as the discrete breather,  are not allowed. 
The absence of the soliton and breather modes in this case is due to the fact that $\Lambda_{2, {\rm soliton}}$ is much smaller than that of $\Lambda_{2, {\rm c}}$ 
and accordingly 
the  $\Lambda_{2, {\rm soliton}}$ (soliton line) lies deep inside the
diffusion region. Due to the exponential dependence of
$\Lambda_{2, {\rm soliton}}$ on $\alpha_0$ (Eq. (31)), it is not
possible to increase $\alpha_0$ till the $\Lambda_{2, {\rm soliton}}$ line
cross the $\Lambda_{2, {\rm c}}$ critical line for the soliton solution to
reappear. Similarly, the discrete breather solutions are also not
allowed in this case since  $\Lambda_{2, {\rm breather}}$  (breather line)
also lies well below the $\Lambda_{2, {\rm c}}$ critical line.\\

\noindent
{\bf 5. Numerical solutions of the higher order discrete nonlinear Schrodinger equation}\\

We have also solved the higher order DNLS equation (Eq. (5)) directly using numerical procedures. This verifies the existence  of the discrete soliton solutions in presence of only three-body interactions, as predicted by the variational analysis. We have  also checked the stability of the soliton solutions over a long time scale. We write $\psi_n(t) = \sqrt {\rho_n (t)} e^{i\theta_n (t)}$ in terms of its two components, as $\psi_n = (x_n, y_n)$ and  $\dot\psi_n = (\dot x_n, \dot y_n)$. Here $x_n$ and $y_n$ denote the real and imaginary parts respectively of $\psi_n$. In terms of these variables, the DNLS (Eq. (5)) can be written as the coupled equations
\begin{equation}
\dot{x_{n}}= - \frac{1}{2}(y_{n-1}+y_{n+1} )+\epsilon_{n}y_{n}+\Lambda_{0} (x^{2}_{n}+y^{2}_{n} )y_n + \Lambda_{2}(x^{2}_{n}+y^{2}_{n})^2 y_n 
\end{equation}
and
\begin{equation}
\dot{y_{n}}=  \frac{1}{2}(x_{n-1}+x_{n+1} )-\epsilon_{n}x_{n}-\Lambda_{0} (x^{2}_{n}+y^{2}_{n} )x_n  - \Lambda_{2}(x^{2}_{n}+y^{2}_{n})^2 x_n 
\end{equation}
Since we want to compare the numerical results with the variational results, we take the initial condition as the variational wavefunction (Eq. (12)) and accordingly write the initial values of $(x_n, y_n)$ as
\begin{equation}
x_{n}=\sqrt{k}e^{-\frac{\left(n-\zeta \right)^{2}}{\alpha}}\cos \left[ p\left(n-\zeta\right)+ \frac{\delta}{2} \left(n- \zeta\right)^{2} \right] 
\end{equation}
and 
\begin{equation}
y_{n}=\sqrt{k}e^{-\frac{\left(n-\zeta \right)^{2}}{\alpha}}\sin \left[ p\left(n-\zeta\right)+ \frac{\delta}{2} \left(n- \zeta\right)^{2} \right]
\end{equation}
where $ \alpha = \gamma^{2}$.
\begin{figure}
\includegraphics[width=200mm,height=200mm]{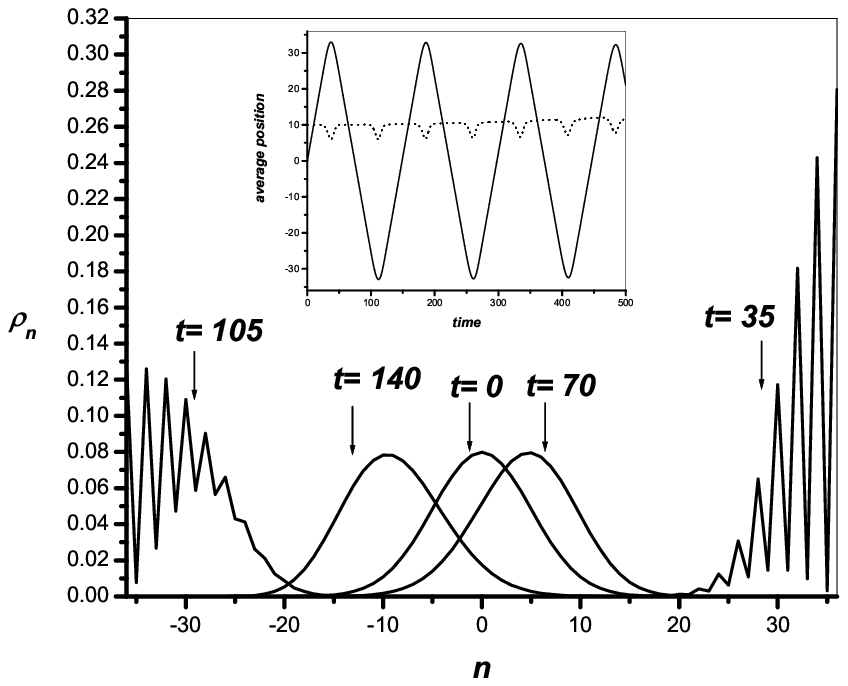}
\caption{ \label{fig.6:}Density profile of the soliton solution in the
  state with  only three-body interaction ($ \Lambda_{2}>0$ and $ \Lambda_{0}=0$) obtained numerically  at different time intervals   $ t= 0, 35, 70, 105$ and $ 140 $ for $\alpha_0=100 $, $p_{0}= 1.6 $ in finite array of 73 sites.  Inset shows the width (dotted line) and average position (solid line) for a much longer time interval.  }
\end{figure}
We solve the coupled nonlinear equations (Eqs.(34, 35))  using the 
Runge-Kutta method, with the initial conditions given by Eqs.(36, 37). The constants of motion, the Hamiltonian (Eq. (11)) and the norm  were checked at every time step of the numerical integration  to verify that these remained constant over time. To connect the numerical results with that of the variational results, we calculate relevant parameters which shows the existence of the soliton solutions. Thus we  calculate the numerical average position which is defined as 
\begin{equation}
<n> =\sum_{n} n \left|\psi_{n}\right|^{2}
\end{equation}
It is easy to show that $<n> =  \zeta(t)$. As mentioned above (Eq. (12)), $\zeta(t)$ is the variational parameter which denote the average position or the center of the density $\rho_n = \mid \psi_n\mid^2$.
We also calculate the numerical width  of the wave packet $ 2\sqrt{<n^{2}>} $ which is defined as
\begin{equation}
<n^{2}>=\sum_{n} n^{2}\left|\psi_{n}\right|^{2} - <n>^{2}
\end{equation}
Again, it can be shown that $<n^{2}> = \gamma^2(t)/4 = \alpha (t)/4$. As mentioned above, $\alpha(t)$ is variational parameter which denote the width of
the wave packet (Eq. (12)). For the soliton solution to exist, the width $<n^{2}>$ should not change with time. Fig.~\ref{fig.6:}  shows a stable  soliton mode collective  excitation  as obtained from the direct numerical integration of the DNLS equation (Eq. (5)) for $\alpha_0=100$. From the figure we can see that the shape of the soliton solution do not change with time. The inset shows the average position and the width over a  long interval of time. This confirms the 
long term stability of the discrete soliton solution.  \\

\noindent
{\bf 6. Conclusion}\\

In conclusion, we have studied BECs in a deep optical lattice with a 
tunable three-body interaction. We have obtained the dynamical phase diagram of the system analytically, using the time-dependent variational method.  The phase diagrams shows  interesting dynamical evolution of the collective excitations  with variation of the strength as well as the sign (attractive or repulsive nature)  of the two- and three-body interactions.  We have explained the causes of the appearance of the phase diagrams of the system. We have also solved the dynamical equation of the system, the higher order discrete nonlinear Schrodinger equation (DNLSE), numerically, to check the existence of the soliton solution in the presence of only three-body interactions, as predicted by the variational analysis.
 As mentioned above, BECs where the three-body interactions dominate over the two-body interactions have been prepared experimentally. It should therefore be possible experimentally to look for the dynamical evolution of the 
collective modes of BECs as discussed above.\\

\noindent
{\bf Acknowledgement} \\

BD thanks SERB, DST, India and BCUD-PU for financial support through  research projects.
 \\

\end{document}